\documentclass[floatfix,amsmath,amssymb,aps,prc,twocolumn,superscriptaddress,showpacs,preprintnumbers,nofootinbib]{revtex4-2}
\usepackage{graphicx,color}
\usepackage{multirow}
\usepackage{times}
\usepackage{bm}
\usepackage{epstopdf}
\usepackage{enumerate}
\usepackage{ulem}
\usepackage{hyperref}
\usepackage{cancel}
\graphicspath{{./Figs/}}

\makeatletter
\def\@fnsymbol#1{\ensuremath{\ifcase#1\or \dagger\or \ddagger\or
   \mathsection\or \mathparagraph\or \|\or **\or \dagger\dagger
   \or \ddagger\ddagger \else\@ctrerr\fi}}
    \makeatother
\newcommand{\deceased}[1]{\altaffiliation{#1}}

\begin{document}
%\preprint{YITP-17-80}

\title{
Femtoscopic study of the $\Lambda\alpha$ interaction
}
\author{Asanosuke Jinno}
\email[]{jinno.asanosuke.36w@st.kyoto-u.ac.jp}
\affiliation{
Department of Physics, Faculty of Science,
Kyoto University, Kyoto 606-8502, Japan}

\author{Yuki Kamiya}
\email[]{kamiya@hiskp.uni-bonn.de}
\affiliation{
Helmholtz Institut für Strahlen- und Kernphysik
and Bethe Center for Theoretical Physics, Universität Bonn, D-53115 Bonn, Germany
}
\affiliation{
RIKEN Interdisciplinary Theoretical and Mathematical Science Program (iTHEMS), Wako 351-0198, Japan
}

\author{Tetsuo Hyodo}
\email[]{hyodo@tmu.ac.jp}
\affiliation{
Department of Physics, Tokyo Metropolitan University, Hachioji 192-0397, Japan
}
\affiliation{
RIKEN Interdisciplinary Theoretical and Mathematical Science Program (iTHEMS), Wako 351-0198, Japan
}

\author{Akira Ohnishi}\deceased{Deceased.}
\affiliation{Yukawa Institute for Theoretical Physics,
Kyoto University, Kyoto 606-8502, Japan}

\date{\today}

\preprint{KUNS-2997}
\begin{abstract}
We examine the $\Lambda$-${}^4\mathrm{He}$ ($\alpha$) momentum correlation in high-energy collisions to elucidate the interaction between Lambdas ($\Lambda$) and nucleons ($N$).
We compare phenomenological $\Lambda\alpha$ potentials with different strengths at short range. In addition to the conventional Gaussian-type potentials, we construct the $\Lambda\alpha$ potentials by substituting the nucleon density distribution in $\alpha$ into the Skyrme-type $\Lambda$ potentials.
We find that the dependence on the employed potential models is visible in the correlation functions from a small-size source.
This indicates that the $\Lambda\alpha$ momentum correlation could constrain the property of the $\Lambda N$ interaction at high densities,
which is expected to play an essential role in dense nuclear matter.
Also, we verify that the Lednicky-Lyuboshits formula can yield erroneous results for a small-size source
with a potential which has a large interaction range, like the $\Lambda\alpha$ system.
\end{abstract}

\maketitle

\section{Introduction}
The interactions between hyperons ($Y$) and nucleons ($N$) are important to investigate various fields of physics,
including hypernuclei, heavy-ion collisions, supernovae, and neutron stars.
For a better understanding of their structures and dynamics, it is crucial to figure out the properties of the $YN$ 
interaction based on the experimental data and astrophysical observations.
The hypernuclear spectroscopy and the $YN$ scattering data have been used for constraining the $YN$ interaction~\cite{Feliciello:2015dua,Gal:2016boi,J-PARCE40:2021qxa,J-PARCE40:2021bgw}.
The observations of neutron stars also impose constraints on the repulsion of the $YN$ interaction through the equation of state (EOS) of the dense nuclear matter.
The precise mass and radius data from NICER~\cite{Miller:2019cac,Miller:2021qha,Riley:2019yda,Riley:2021pdl}
and the gravitational wave data from LIGO and Virgo~\cite{LIGOScientific:2017vwq}
have been utilized to constrain the stiffness of the EOS.

Although more and more constraints on the $YN$ interactions have been imposed, a problem with the appearance of hyperons in neutron stars (hyperon puzzle) is still intensively discussed~\cite{Gal:2016boi,Burgio:2021vgk}.
The hyperon puzzle refers to the problem that most EOSs including hyperons become too soft to support massive neutron stars as heavy as twice the solar mass~\cite{Demorest:2010bx,Antoniadis:2013pzd,Fonseca:2016tux,NANOGrav:2019jur, Miller:2021qha}.
Many solutions to the puzzle have been proposed,
such as repulsive $YY$ interactions ~\cite{Weissenborn:2011ut,Chatterjee:2015pua, Fortin:2017cvt} and/or many-body baryon interactions~\cite{Nishizaki:2002ih, Lonardoni:2014bwa, Nagels:2015lfa, Gerstung:2020ktv, Yamamoto:2013ada, Yamamoto:2014jga, Nagels:2015lfa, Yamamoto:2015lwa, Togashi:2016fky, Haidenbauer:2016vfq, Logoteta:2019utx} at high densities,
and the crossover transition from hadronic matter to quark matter before the appearance of hyperons~\cite{Baym:2017whm, Kojo:2021wax}.
However, a definitive conclusion is not obtained on whether hyperons can appear in neutron stars,
due to the lack of constraints on the interaction between baryons at high densities based on the experimental data.

Studying the interaction between the Lambda hyperon ($\Lambda$) and $^4\mathrm{He}$ ($\alpha$) would be another approach to tackle the hyperon puzzle. Because the central density in $\alpha$ reaches about twice the normal nuclear density~\cite{Hofstadter:1957wk}, the short range part of the $\Lambda\alpha$ interaction may reflect the behavior of the $\Lambda N$ interaction at high densities.
Many models of the $\Lambda\alpha$ interaction have been constructed by folding the $\Lambda N$ interaction, and they have been applied in the few-body calculations of the $\Lambda$ hypernuclei~\cite{Motoba:1985uj,Hiyama:1997ub,Kanada-Enyo:2017ynk,Myo:2023ipo}.
In the construction of the $\Lambda\alpha$ interaction, the strength of the total attraction is determined by
the $\Lambda$ binding energy of $^5_\Lambda\mathrm{He}$
\cite{Juric:1973zq} and
the approximate interaction range of the $\Lambda\alpha$ system can be inferred from the meson exchange picture.
Then, the missing piece of the information on the $\Lambda \alpha$ interaction is its behavior at short range,
which is expected to reflect the $\Lambda N$ interaction in dense nuclear matter due to the central density in $\alpha$.
The $\Lambda \alpha$ interaction at short range has been studied by the weak decay of the $\Lambda$ hypernuclei~\cite{Motoba:1991uv,Kumagai-Fuse:1994ulj,Botta:2012xi}.
According to Ref.~\cite{Kumagai-Fuse:1994ulj}, the mesonic weak-decay widths of the light $\Lambda$ hypernuclei are better reproduced by using the $\Lambda\alpha$ potential with stronger central repulsion.
The origin of the central repulsion is, however, not well understood, and the microscopic understanding of the repulsive core is highly desired.

Recently, the momentum correlation functions measured in high-energy collisions have been extensively employed
for investigating the baryon-baryon interactions which are difficult to measure in usual scattering experiments,
such as
$p\Lambda$~\cite{STAR:2005rpl_pLam,ALICE:2018ysd_NL_LL,ALICE:2021njx_pLpS},
$p\Sigma^0$~\cite{ALICE:2019buq_NS0},
$\Lambda \Lambda$~\cite{STAR:2014dcy,ALICE:2018ysd_NL_LL,ALICE:2019eol,Isshiki:2021bqh},
$p\Omega^{-}$~\cite{STAR:2018uho,ALICE:2020mfd},
$p \Xi^{-}$~\cite{ALICE:2019hdt,ALICE:2020mfd,Isshiki:2021bqh},
$\Lambda \Xi$~\cite{ALICE:2022uso}, and $\Xi\Xi$~\cite{Isshiki:2021bqh}
systems.
When the interaction range of the two-body system is much shorter than the spatial size of the particle emission source, the correlation function can be obtained by
the Lednicky-Lyuboshits (LL) formula~\cite{Lednicky:1981su} expressed by the low-energy scattering parameters. In fact, the LL formula has been applied in the analysis of the correlation functions from high-energy collisions~\cite{STAR:2005rpl_pLam,STAR:2014dcy,ALICE:2020wvi,ALICE:2021szj}.
Furthermore, the momentum correlation for the mass number $A\ge3$ systems comes to be studied to access the many-body interactions
both experimentally
($p$-deuteron ($d$)~\cite{Singh:2022qmg_ALICEpd},
and $pp\Lambda$~\cite{ALICE:2022boj_ppL})
and theoretically 
($pd$~\cite{Viviani:2023kxw},
$\Lambda d$~\cite{Haidenbauer:2020uew},
and $\Xi d$~\cite{Ogata:2021mbo}).

The momentum correlation between $\Lambda$ and $\alpha$ would be a promising probe for the $\Lambda \alpha$ interaction.
Since the scattering length and the effective range are empirically fixed for the $\Lambda \alpha$ system to some extent,
more detailed information on the potential shape could be accessed in the femtoscopic study.
Since the total spin of $\alpha$ is zero, the $\Lambda\alpha$ interaction can be directly accessed from the measured spin-summed correlation function.
This is in sharp contrast, for instance, to the $\Lambda N$ interaction where the different spin components contribute to the measured correlation function and the separation of the components is not straightforward.
Also, because the range of the $\Lambda\alpha$ interaction is longer than the baryon-baryon interactions, the LL formula may not be applicable for small size sources.
In such cases, the momentum correlation functions reflect the behavior of the scattering wave function in the interaction region.

In this paper, we give predictions on the 
$\Lambda\alpha$ momentum correlation functions with five different $\Lambda\alpha$ potential models. The Isle and single-range Gaussian (SG) models are developed phenomenologically~\cite{Kumagai-Fuse:1994ulj}.
We newly construct the LY-IV, Chi3, and Chi3 w/o mom models by substituting the density distribution in $\alpha$ into the Skyrme-type $\Lambda$ potentials~\cite{Lanskoy:1997xq,Jinno:2023xjr} in order to relate the in-medium properties of $\Lambda$ and the $\Lambda\alpha$ potential.
While the binding energy of $^{5}_\Lambda\mathrm{He}$ is fixed to the empirical value, those interaction models have different properties at short range.
We first demonstrate that the difference of the $\Lambda\alpha$ potentials can be seen in the correlation functions from a small size source. We also verify the LL formula gives the incorrect estimation for the correlation from the small sources.
The comparison with future experimental data will lead to
the constraints on the $\Lambda N$ interaction in nuclear matter.

This paper is organized as follows.
In Sec.~\ref{sec:Methods}, we introduce five models of the $\Lambda\alpha$ potential
and discuss their properties by focusing on the short-range behavior. We also present the methods to calculate the correlation function of the $\Lambda\alpha$ system.
In Sec.~\ref{sec:CorrFunc}, we show the results of the calculated $\Lambda\alpha$ correlation function
and explore the conditions under which the differences among the models are enhanced.
The conclusions and outlook are given in Sec.~\ref{sec:summary}.
Throughout the text, we work in the natural unit $\hbar=c=1$.

\section{Methods}
\label{sec:Methods}

\subsection{Formulation of $\Lambda\alpha$ potentials}

First, we introduce two phenomenological $\Lambda\alpha$ potential models, Isle and SG~\cite{Kumagai-Fuse:1994ulj}.
They are Gaussian-type potentials expressed as
\begin{equation}
\label{eq:Gauss}
    U_{\Lambda \alpha} =
    \sum_i U_i \exp\left[{-\left(\dfrac{r}{a_i}\right)^2}\right],
\end{equation}
where $U_i$ and $a_i$ are the potential parameters.
Both models are fitted to the experimental binding energy of
$^{5}_\Lambda\mathrm{He}$~\cite{Juric:1973zq}.
The Isle potential has a two-range Gaussian form with
central repulsion and long-range attraction.
The SG potential has a single attractive Gaussian form and has the same behavior with a folding $\Lambda\alpha$ potential using a one-range Gaussian potential~\cite{Motoba:1985uj}, which is used for few-body hypernuclear calculations~\cite{Motoba:1985uj,Myo:2023ipo}.
The parameters of the Isle and SG potentials are listed in Table~\ref{tab:GaussParams}.

\begin{table}[tbp]
\caption{Parameters of the Gaussian-type $\Lambda\alpha$ potential in Eq.~\eqref{eq:Gauss}.}
\centering
%\begin{tabular}{l@{\qquad}|rr}
\begin{tabular}{c|cc}
\hline
\hline
& Isle & SG \\
\hline
$U_1~\textrm{(MeV)}$ & $\phantom{-}450.4$ & $-43.92$\\
$U_2~\textrm{(MeV)}$ & $-404.9$ & $0.0$\\
$a_1~\textrm{(fm)}$ & $1.25$ & $1.566$\\
$a_2~\textrm{(fm)}$ & $1.41$ & -\\
\hline
\hline
\end{tabular}
\label{tab:GaussParams}
\end{table}

Next, we construct $\Lambda\alpha$ potential models based on the
Skyrme-type $\Lambda$ potentials in nuclear matter, Chi3~\cite{Jinno:2023xjr}
and LY-IV~\cite{Lanskoy:1997xq}.
The Chi3 potential is constructed by reproducing the results of the density dependence~\cite{Gerstung:2020ktv} and the momentum dependence~\cite{Kohno:2018gby} of the $\Lambda$ potential from chiral effective field theory with $\Lambda NN$ three-body interaction and $\Lambda NN$-$\Sigma NN$ coupling.
The LY-IV potential is constructed from the results of the $\Lambda$ potential in nuclear matter by the $G$-matrix calculation of the Nijmegen model F~\cite{Nagels:1976xq,Nagels:1978sc}.
Although both the Chi3 and LY-IV potentials reproduce the binding energies of the hypernuclei from $^{13}_\Lambda \mathrm{C}$ to $^{208}_\Lambda \mathrm{Pb}$~\cite{Jinno:2023xjr}, they exhibit different behavior at high densities.
The Chi3 potential is so repulsive at high densities
that it avoids the appearance of the hyperons in the neutron stars~\cite{Gerstung:2020ktv}.
On the other hand, the attractive nature of the LY-IV potential at high densities allows the appearance of hyperons in neutron stars.
Therefore, distinguishing between Chi3 and LY-IV is important to advance the discussion on the hyperon puzzle of neutron stars.

We construct the $\Lambda\alpha$ potential from the Skyrme-type $\Lambda$ potential in nuclear matter.
For $\Lambda$ hypernuclei with the baryon number $A$, the Skyrme-Hartree-Fock equation for the single-particle wave function of the $i$th baryon $B_i$ is given as~\cite{Rayet:1976fs,Rayet:1981uj}
\begin{multline}
\label{eq:SHFeq}
    \Bigg[ - \nabla \cdot \left(\dfrac{1}{2 m_{B_i}^*(\bm{r})} \nabla\right) + U_{B_i}(\bm{r}) \\
    - i \bm{W}_{B_i}(\bm{r}) \cdot (\nabla \times {\bm{\sigma}}) \Bigg] \phi_i = \epsilon_i \phi_i,
\end{multline}
where $\phi_i$, $\epsilon_i$, and $\bm{\sigma}$ are the single-particle wave function, the single-particle energy, and the Pauli matrices acting on the spin wave function, respectively.
The effective mass $m_{B_i}^*$ and the single-particle potential $U_{B_i}$ for $B_i=\Lambda$ are defined as~\cite{Jinno:2023xjr}
\begin{align}
    \dfrac{1}{2m_\Lambda^*(\bm{r})} &= \dfrac{1}{2 m_\Lambda} + a^\Lambda_2 \rho_N(\bm{r}), 
    \label{eq:mLambdastar}\\
    U_\Lambda(\bm{r}) &=  a^\Lambda_1\rho_N(\bm{r}) + a^\Lambda_2 \tau_N(\bm{r}) - a^\Lambda_3 \Delta \rho_N(\bm{r}) \nonumber \\
    &\quad + a^\Lambda_4 \rho^{4/3}_N(\bm{r}) + a^\Lambda_5 \rho^{5/3}_N(\bm{r}), \label{eq:ULambda}
\end{align}
respectively, where $a^\Lambda_j$ ($j=1,\dotsb, 5$) are the parameters of the Skryme potential.
The nucleon density $\rho_N$ and the kinetic density $\tau_N$ are defined as
\begin{align}
    \label{eq:densities}
    \rho_N = \sum_{i=1}^{A-1} |\phi_i|^2,~~~\tau_N = \sum_{i=1}^{A-1} |{\nabla \phi_i}|^2,
\end{align}
respectively where the index $i=A$ corresponds to $\Lambda$.
In the following, we ignore the spin-orbit potential $\bm{W}_{B_i}$ for $B_i=\Lambda$ because it is expected to be small from the experimental data~\cite{Ajimura:2001na,Akikawa:2002tm}.

The $\Lambda\alpha$ potential $U_{\Lambda \alpha}$ is constructed by substituting
the $\alpha$-particle density distribution in the Skyrme-type $\Lambda$ potential~\eqref{eq:ULambda}.
The wave function in the $\alpha$ particle is assumed by
a product of Gaussians with the width parameter $\nu$ as~(e.g., Ref.~\cite{Akaishi:1986gm}) follows:
\begin{align}
    \label{eq:AlphaWF}
    \psi(\bm{r}_1,\cdots,\bm{r}_4)&=\prod^{4}_{i=1} \phi(\bm{r}_i), \\
    \phi(\bm{r}) &= \left(\dfrac{2\nu}{\pi}\right)^{3/4} \exp[-\nu \bm{r}^2].
\end{align}
With this wave function, the nucleon density and the kinetic density measured from the center-of-mass $\bm{r}_G = \sum^{4}_i \bm{r}_i/4$
are given as
\begin{align}
    \rho_N(\bm{r})&=
    \int d \bm{r}_1 \cdots d\bm{r}_4 
    \left| \psi(\bm{r}_1, \cdots, \bm{r}_4) \right|^2 
    \sum^4_{i=1} \delta^3 (\bm{r}_i - \bm{r}_G - \bm{r}) \nonumber \\
    &= 4\left(\dfrac{2 \nu_c}{\pi}\right)^{3/2}
    \exp [-2\nu_c \bm{r}^2],~\text{where}~~\nu_c = \dfrac{4}{3}\nu,
    \label{eq:densityCM}
    \\
    \tau_N(\bm{r})&=
    \int d \bm{r}_1 \cdots d\bm{r}_4 \nonumber \\
    &\quad \times\sum^4_{i=1} \delta^3 (\bm{r}_i - \bm{r}_G - \bm{r})
    \left| \nabla_i \psi(\bm{r}_1, \cdots, \bm{r}_4) \right|^2  \nonumber \\
    &= \rho_N(\bm{r}) \left(4 \nu^2 \bm{r}^2 + \dfrac{3}{4}\nu\right).
    \label{eq:KindensityCM}
\end{align}

Now, the Skyrme-type $\Lambda\alpha$ potential~\eqref{eq:ULambda} with densities~\eqref{eq:densityCM} and \eqref{eq:KindensityCM} has six parameters:
the Skryme parameters $a^\Lambda_j$ ($j=1,\dotsb,5$) and the width parameter of the $\alpha$ wave function $\nu$.
To respect the property of the $\Lambda$ potential in nuclear matter,
we fix $a^\Lambda_1$, $a^\Lambda_2$, $a^\Lambda_4$, and $a^\Lambda_5$
by the values in the previous study of the $\Lambda$ hypernuclei~\cite{Jinno:2023xjr}.
In this study, we adopt the Chi3 and LY-IV models in Ref.~\cite{Jinno:2023xjr}.
The width parameter $\nu$ is determined to reproduce
the root-mean-square (rms) charge radius of the $\alpha$ particle.
The rms charge radii of the $\alpha$ particle is calculated as
\begin{align}
    \langle \bm{r}^2 \rangle_\alpha =
    \dfrac{1}{4} \sum^4_{i=1} \langle (\bm{r}_i - \bm{r}_G)^2 \rangle
    = \dfrac{9}{16 \nu}.
\end{align}
With this relation,
$\nu$ is determined as $0.20~\text{fm}^{-2}$
to reproduce the experimental value of the $\alpha$ charge radius $\sqrt{\langle \bm{r}^2 \rangle_A} = 1.67824(83)~\text{fm}$~\cite{Krauth:2021foz}.
We note that the proton charge radius is not 
included here, which is usually taken into account~\cite{Etminan:2019gds,Akaishi:1986gm}.
If the proton charge were included, the range of the $\Lambda\alpha$ potential would become
too short because of the zero range nature of 
the Skyrme interaction.
Finally, the parameter $a^\Lambda_3$ is determined to reproduce the experimental data of the $\Lambda$ binding energy
of ${}^5_\Lambda \mathrm{He}$ of $3.12~\text{MeV}$~\cite{Juric:1973zq} in each model.
The parameters of the Skyrme-type $\Lambda\alpha$ potentials Chi3 and LY-IV are listed in Table~\ref{tab:SkyrmeParams}.

There is another way to fix the parameters of $a^\Lambda_3$ and $\nu$; fix $a^\Lambda_3$ to reproduce the $\Lambda$ binding energy of ${}^{13}_\Lambda \mathrm{C}$ and then determine $\nu$ to reproduce the $\Lambda$ binding energy of ${}^5_\Lambda \mathrm{He}$.
In this case, $\nu$ is obtained as $0.19~\text{fm}^{-2}$ for Chi3.
Because this method gives essentially the same parameters, here we take the first approach to determine $a^\Lambda_3$ and $\nu$.

\begin{table}[tbp]
\caption{Sets of parameters for the Skyrme-type $\Lambda\alpha$
potential~\eqref{eq:ULambda}.
We note that the parameters $a_3^\Lambda$ are modified from the values in Refs.~\cite{Lanskoy:1997xq,Jinno:2023xjr} to reproduce the $^{5}_\Lambda \mathrm{He}$ binding energy.
}
\centering
\begin{tabular}{l@{\qquad}rrr}
\hline
\hline
& Chi3 & LY-IV & Chi3 w/o mom\\
\hline
$a_1^\Lambda~({\text{MeV}}~{\text{fm}^3})$ & $-388.3$ & $-500.9$ & $-388.3$\\
$a_2^\Lambda~({\text{MeV}}~{\text{fm}^5})$ & $47.3$ & $16.0$ & $0$\\
$a_3^\Lambda~({\text{MeV}}~{\text{fm}^5})$ & $30.8$ & $13.5$ & $17.9$\\
$a_4^\Lambda~({\text{MeV}}~{\text{fm}^4})$ & $-405.7$ & $548.4$ & $-405.7$\\
$a_5^\Lambda~({\text{MeV}}~{\text{fm}^5})$ & $1257$ & $0$ & $1428$\\
$\nu~({\text{fm}^{-2}})$ & $0.20$ & $0.20$ & $0.20$\\
\hline
\hline
\end{tabular}
\label{tab:SkyrmeParams}
\end{table}

The Skyrme-type $\Lambda\alpha$ potentials reflect the momentum-dependent part of the $\Lambda$ potentials which are not well constrained by the experimental data.
It is therefore desirable to examine the effect of the momentum dependence of the potential in the correlation functions. To this end, we switch off the momentum dependence in the Chi3 model by setting $a^\Lambda_2=0$.
To keep the $\Lambda$ potential in symmetric nuclear matter unchanged,
we modify the value of $a_5^\Lambda$.
In the Fermi gas model of the symmetric nuclear matter with zero temperature, the single-particle wave function is the plane wave, and the energy density level is filled up to the Fermi momentum $k_F=(3\pi\rho_N/2)^{1/3}$. Then, $\tau_N$~\eqref{eq:densities} becomes
\begin{align}
    \tau_N &= g_N \int_{|\bm{k}|\leq k_F}\dfrac{d^3 k}{(2\pi)^3} |\nabla e^{i\bm{k}\cdot\bm{x}}|^2 \\
    &= \dfrac{3}{5}\left(\dfrac{3\pi^2}{2}\right)^{2/3} \rho_N^{5/3},
\end{align}
where $g_N=4$ is the spin and isospin degeneracy.
Therefore, we increase $a^\Lambda_5$ by $47.3~\text{MeV}~\text{fm}^5\times3/5(3\pi^2/2)^{2/3}$,
which compensates the contribution from the $a_2^\Lambda$ term in the original model.
The width parameter $\nu$ and $a^\Lambda_3$ are determined
by the same procedure as Chi3.
We call this $\Lambda\alpha$ potential Chi3 w/o mom, whose parameters are also summarized in Table~\ref{tab:SkyrmeParams}.

\subsection{Properties of $\Lambda\alpha$ potentials}

\begin{figure}[tbp]
\includegraphics[width=8cm]{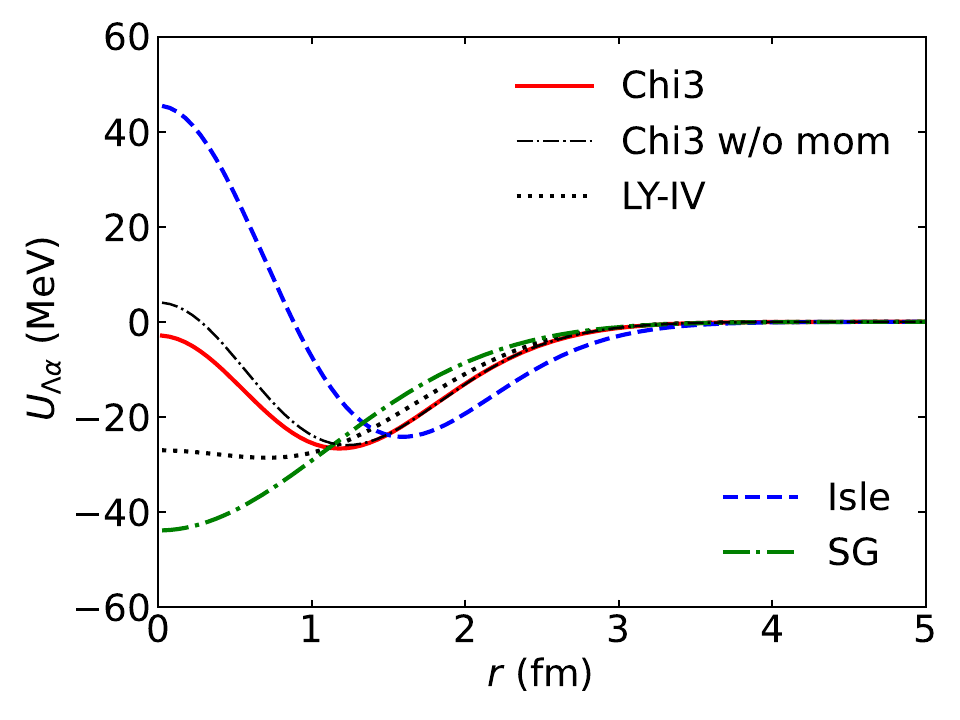}
\caption{$\Lambda\alpha$ potentials as functions of the distance between $\Lambda$ and $\alpha$.
Isle (dashed line) and SG (thick dash-dotted line) are the phenomenological potentials given in Gaussian form~\cite{Kumagai-Fuse:1994ulj}.
Chi3 (solid line), LY-IV (dotted line), and Chi3 w/o mom (thin dash-dotted line) are the Skyrme-type $\Lambda$ potentials with 
the $\alpha$ density distribution.
}
\label{fig:LamAlphaPot}
\end{figure}

In Fig.~\ref{fig:LamAlphaPot}, we show the Skyrme type $\Lambda\alpha$ potentials, Chi3, LY-IV, and Chi3 w/o mom and the Gaussian-type SG and Isle.
The main difference among the models is in the behavior at short distance. Among the Skyrme-type potentials, the Chi3 and Chi3 w/o mom potentials increase near the origin, while LY-IV exhibits the Woods-Saxon like shape. This is a consequence of the different high-density behavior of the $\Lambda$ potential in nuclear matter mentioned above. In this way, we explicitly show that the property of $\Lambda$ in nuclear matter is reflected in the short-range behavior of the $\Lambda\alpha$ potential. The Isle potential has a further strong repulsive core at a short distance, while the SG model is entirely attractive. In all cases, the interaction ranges are of the order of $2$-$3~\text{fm}$.

\begin{figure}[tbp]
\includegraphics[width=8cm]{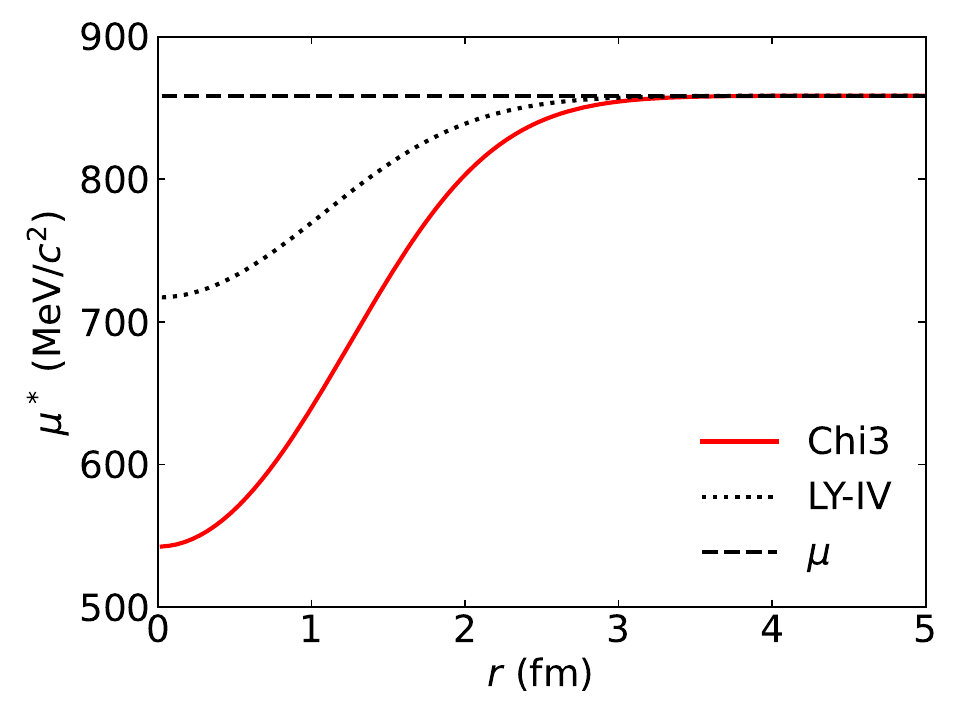}
\caption{
Reduced effective masses as functions
of the distance between $\Lambda$ and $\alpha$ for Chi3 (solid line) and LY-IV (dotted line).
Its vacuum value~$\mu$ corresponds to the dashed line.
}
\label{fig:EffMass}
\end{figure}

The two-body Schr\"odinger equation for the $\Lambda\alpha$ system is written as
\begin{multline}
\label{eq:SchEq}
    \Bigg[-\nabla_\Lambda \cdot \left(\dfrac{1}{2m^*_\Lambda(\bm{r})} \nabla_\Lambda \right)
    - \dfrac{1}{2m_\alpha} \nabla_\alpha^2 \\
    + U_{\Lambda \alpha}(\bm{r})\Bigg] \Phi(\bm{r}_\Lambda,\bm{r}_\alpha) = E \Phi(\bm{r}_\Lambda,\bm{r}_\alpha),
\end{multline}
where $\bm{r}_i$ is the coordinate of the particle $i$.
The derivative operator $\nabla_i$ is acting on the particle $i$ and 
the relative coordinate is defined as $\bm{r}=\bm{r}_\alpha-\bm{r}_\Lambda$.
The effective mass $m^*_\Lambda(\bm{r})$ of $\Lambda$ is set as its vacuum value $m_\Lambda$ for local potentials: Isle, SG, and Chi3 w/o mom.
In the center-of-mass frame, the total momentum is zero, and then $\nabla_{\bm{R}} \Phi=\bm{0}$ with the center-of-mass coordinate $\bm{R}=(m_\alpha \bm{r}_\alpha + m_\Lambda \bm{r}_\Lambda)/(m_\alpha + m_\Lambda)$, and the Schr\"odinger equation~\eqref{eq:SchEq} can be reduced to the equation for the relative wave function $\Psi$ as
\begin{equation}
\label{eq:RelSchEq}
    \left[-\nabla_{\bm{r}} \cdot \left(\dfrac{1}{2\mu^*(\bm{r})} \nabla_{\bm{r}}\right) + U_{\Lambda \alpha}(\bm{r})\right]
    \Psi(\bm{r}) = E \Psi(\bm{r}),
\end{equation}
where we call $\mu^* = m^*_\Lambda m_\alpha/(m^*_\Lambda + m_\alpha)$ the reduced effective mass.
In Fig.~\ref{fig:EffMass}, the $r$ dependence of $\mu^*$ for different models is shown.
The reduced effective mass is a constant $\mu=m_\Lambda m_\alpha/(m_\Lambda + m_\alpha)$ for local potentials, Isle, SG, and Chi3 w/o mom. For nonlocal potentials, the reduced effective mass decreases from $\mu$ in the distance where the nucleon density appears, and Chi3 shows stronger reduction than that of LY-IV. The reduction of $\mu^*$ is a consequence of positive $a_2^\Lambda$ [see Eq.~\eqref{eq:mLambdastar}], which is enhanced for the model with larger $a_2^\Lambda$.

\begin{figure}[tbp]
\includegraphics[width=8cm]{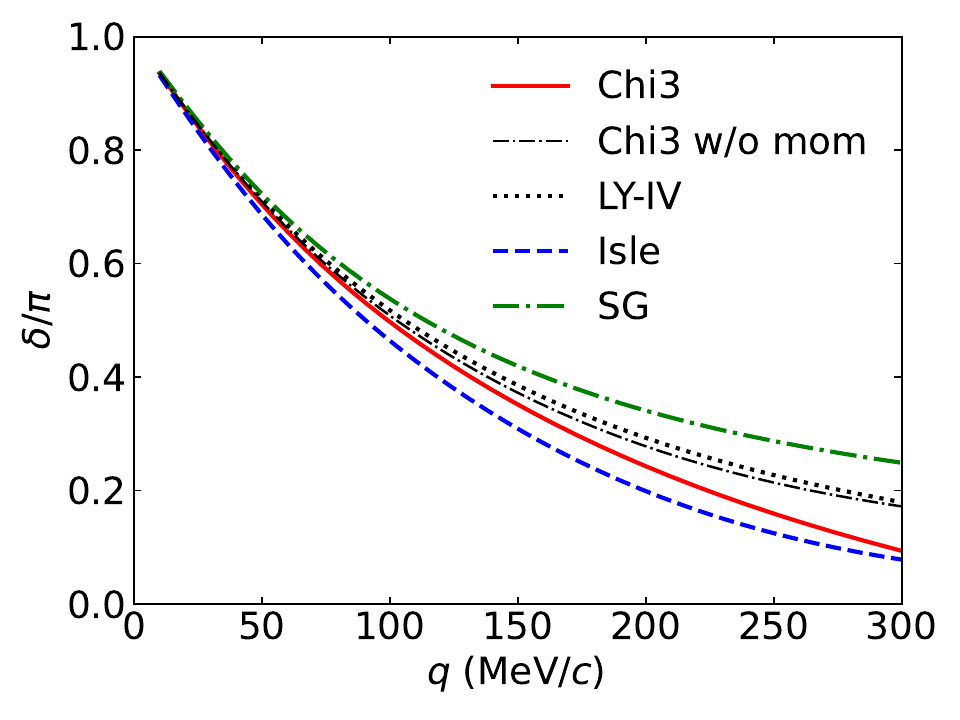}
\caption{Normalized $\Lambda\alpha$
scattering phase shift $\delta/\pi$ obtained from the relative Schr\"odinger equation~\eqref{eq:RelSchEq} with five $\Lambda\alpha$ potential models: Chi3 (solid line), Chi3 w/o mom (thin dash-dotted line), LY-IV (dotted line), Isle (dashed line), and SG (thick dash-dotted line).
}
\label{fig:PhaseShift}
\end{figure}

In Fig.~\ref{fig:PhaseShift}, normalized $\Lambda\alpha$ phase shifts $\delta/\pi$ calculated with various potential models are shown as functions of the magnitude of the relative momentum $q=\sqrt{2\mu E}$.\footnote{To determine the momentum, we use the reduced mass $\mu$ also for the nonlocal potentials, because the scattering momentum is defined in the asymptotic region $r\to \infty$ where $\mu^*\to\mu$.} The behavior of the low-energy phase shift is constrained by the bound state $^{5}_\Lambda \mathrm{He}$ below the threshold.
The $\Lambda$ binding energy of $^{5}_\Lambda \mathrm{He}$ is listed in Table~\ref{tab:Properties}.
The results are similar since all models are constructed to reproduce the experimental value.
The scattering length $a_0$ and the effective length $r_{\rm eff}$ are defined with the effective-range expansion parameters as
\begin{equation}
    q \cot\delta=-\dfrac{1}{a_0}+\dfrac{1}{2}r_{\rm eff}q^2+\mathcal{O}\left(q^4\right).
    \label{eq:ERE}
\end{equation}
Obtained values are listed in Table~\ref{tab:Properties}.
We note that the ordering of the magnitude of $a_0$ and $r_{\rm eff}$ coincides with the ordering of the value of the potential $U_{\Lambda\alpha}$ at $r=0$,
except for Chi3 w/o mom.
To check the convergence of the effective-range expansion, we evaluate the binding energy estimated by the truncated effective-range expansion~\cite{Hyodo:2013iga},
\begin{equation}
    B_\Lambda^{\rm ERE}=-\frac{1}{2\mu}
    \left(\frac{i}{r_{\rm eff}}-\frac{1}{r_{\rm eff}}\sqrt{\frac{2r_{\rm eff}}{a_0}-1}\right)^2 ,
    \label{eq:BLambdaERE}
\end{equation}
in Table~\ref{tab:Properties}. It is seen that the exact binding energy $B_\Lambda$ is reasonably estimated by $B_\Lambda^{\rm ERE}$, indicating the good convergence of the effective-range expansion. At the same time, however, the deviation of $B_\Lambda$ and $B_\Lambda^{\rm ERE}$ increases for models with larger $r_{\rm eff}$.

\begin{table}[tbp]
\caption{$\Lambda$ binding energy of ${}^5_\Lambda \mathrm{He}$, scattering length $a_0$,
and effective range $r_{\rm eff}$ for five $\Lambda\alpha$
potential models. $B_\Lambda^{\rm ERE}$ represents the binding energy estimated by the effective-range expansion~\eqref{eq:BLambdaERE}.}
\centering
\begin{tabular}{l@{\qquad}rrrrr}
\hline
\hline
& Isle & SG & Chi3 & LY-IV & Chi3 w/o mom\\
\hline
$B_\Lambda~({\text{MeV}})$ & $-3.10$ & $-3.09$ & $-3.12$ & $-3.12$ & $-3.12$\\
$a_0~({\text{fm}})$ & $4.24$ & $3.79$ & $4.01$ & $3.89$ & $3.95$\\
$r_{\rm eff}~({\text{fm}})$ & $2.07$ & $1.56$ & $1.84$ & $1.70$ & $1.77$\\
$B_\Lambda^{\rm ERE}~({\text{MeV}})$ & $-3.79$ & $-3.13$ & $-3.41$ & $-3.26$ & $-3.33$ \\
\hline
\hline
\end{tabular}
\label{tab:Properties}
\end{table}

\subsection{Correlation function}
To calculate the $\Lambda\alpha$ correlation function $C(q)$,
we employ the Koonin-Pratt (KP) formula~\cite{Koonin:1977fh,Pratt:1986cc,ExHIC:2017smd}
\begin{equation}
    C(q)=\int d\bm{r} S(\bm{r}) \left|\Psi^{(-)}(\bm{r},\bm{q})\right|^{2},
\end{equation}
where we assume the source function $S$ as a spherical and static Gaussian source function
$S(r)=\exp(-r^2/4R^2)/(4\pi R^2)^{3/2}$
with the source size $R$.
For the relative wave function $\Psi^{(-)}$ with the outgoing boundary condition,
we include the interaction effect only in the $s$-wave state since the sizable correlation emerges
only in the low-momentum region where the $s$-wave scattering dominates.
Then, the relative wave function is written as
\begin{equation}
    \Psi^{(-)}(\bm{r},\bm{q}) = \exp\left(i\bm{r}\cdot\bm{q}\right)-j_0(qr)+\chi_q(r),
\end{equation}
where $j_0=\sin(qr)/qr$ and $\chi_q$ is the $s$-wave radial wave function
calculated from Eq.~\eqref{eq:RelSchEq}.
The resulting correlation function is expressed as
\begin{multline}
    C(q)=1+ \int d\bm{r} S(r)\left[|\chi_q(r)|^2-\left(j_0(qr)\right)^2\right]. \label{eq:KPRed}
\end{multline}

To express the correlation function with the scattering length and the effective range, 
the Lednicky-Lyuboshits (LL) formula~\cite{Lednicky:1981su}
\begin{align}
    C_{\mathrm{LL}}(q)
    &=1
    +\dfrac{|f(q)|^{2}}{2R^2}F_3\left(\dfrac{r_{\rm eff}}{R}\right) 
    \nonumber \\
    &\quad +\dfrac{2\mathrm{Re}f(q)}{\sqrt{\pi}R} F_1 (2qR)
    -\dfrac{\mathrm{Im}f(q)}{R} F_2 (2qR) \label{eq:LL}
\end{align}
has been utilized in various hadron-hadron systems.
The scattering amplitude $f=(q\cot{\delta}-iq)^{-1}$ is calculated with Eq.~\eqref{eq:ERE} by neglecting $\mathcal{O}(q^4)$ terms.
The functions $F_i$ are defined as
$F_1(x)=\int^x_0 dt e^{t^2-x^2}/x$,
$F_2(x)=(1-e^{-x^2})/x$,
and $F_3(x)=1-x/(2\sqrt{\pi})$.
In deriving the LL formula, the wave function $\chi_q(r)$ in Eq.~\eqref{eq:KPRed} is approximated by the asymptotic wave function for the whole radial range.
This approximation works for the case when
the source size is much larger than the interaction range.
However, compared to the typical source size used for the baryon-baryon femtoscopy, which is $1$-$5~\rm{fm}$ depending on the collision conditions,
the interaction range of $U_{\Lambda\alpha}$ can be comparable or even larger.
In the next section, we check the validity of the LL formula for the various source sizes.

\section{$\Lambda\alpha$ correlation function}
\label{sec:CorrFunc}
\begin{figure*}[tbp]
\includegraphics[width=5.9cm]{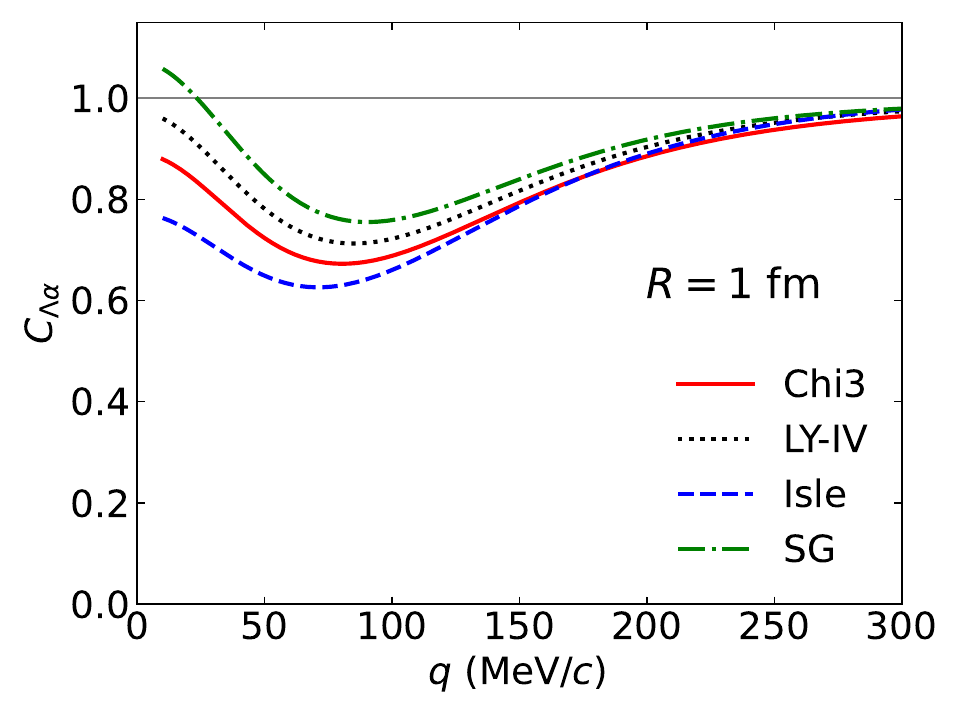}
\includegraphics[width=5.9cm]{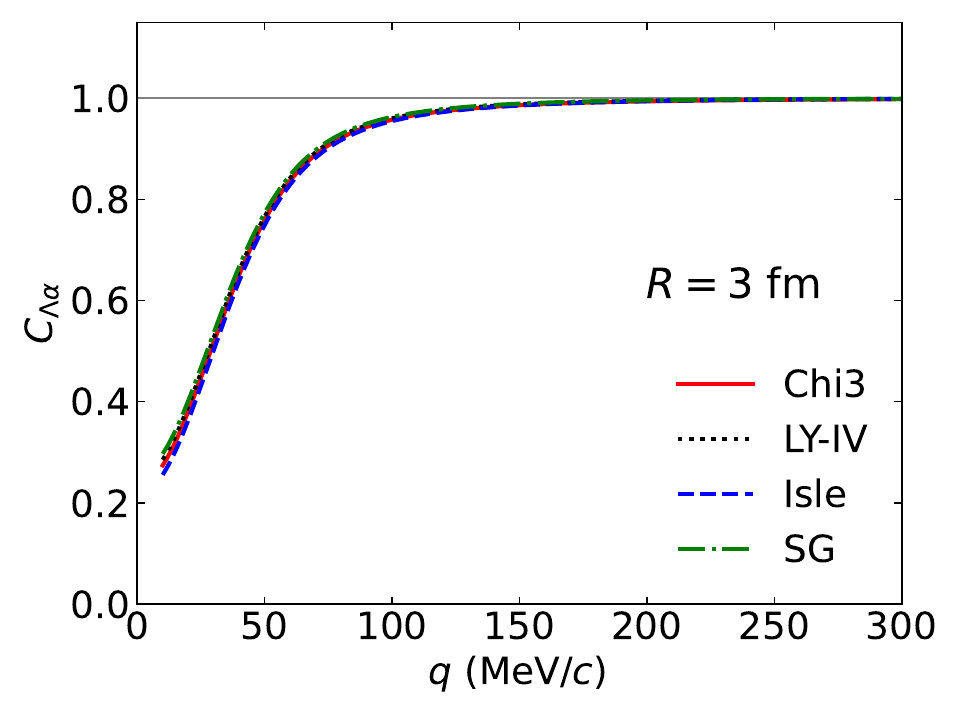}
\includegraphics[width=5.9cm]{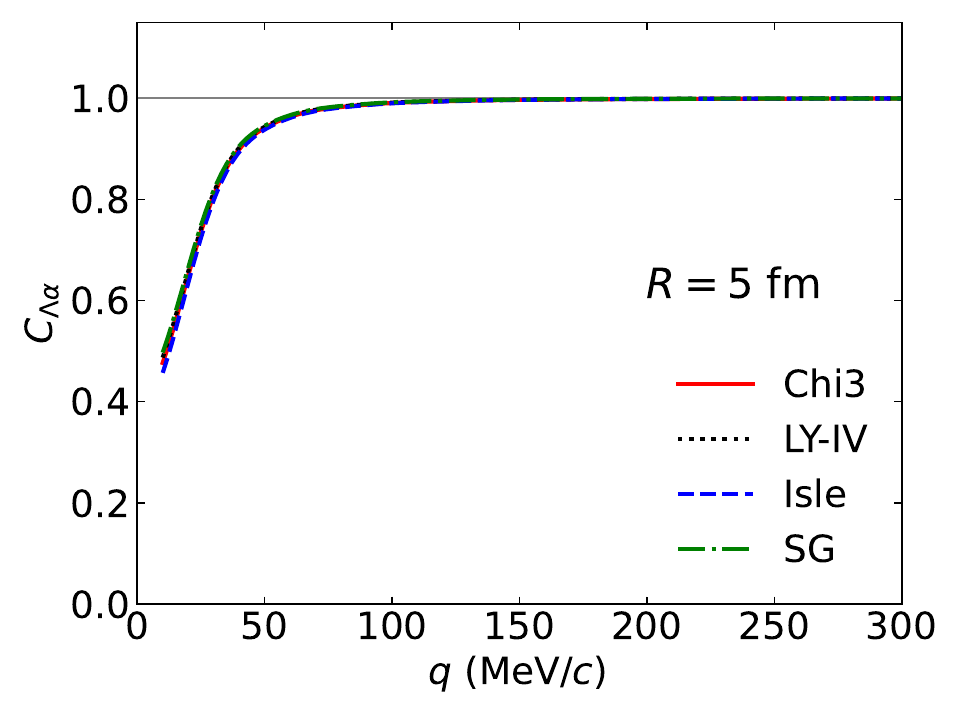}
\caption{$\Lambda\alpha$ correlation functions for three different source sizes.
The solid and dotted lines show the result calculated
by the Skyrme-type $\Lambda\alpha$ potentials, Chi3 and LY-IV, respectively.
The dashed and dash-dotted lines are the results from
the phenomenological $\Lambda\alpha$ potentials, Isle and SG, respectively.
}
\label{fig:CF_R1}
\end{figure*}

In this section, we present the numerical results of the $\Lambda\alpha$ correlation functions $C_{\Lambda\alpha}$.
In Fig.~\ref{fig:CF_R1}, we show the $\Lambda\alpha$ correlation functions calculated with various $\Lambda\alpha$ potential models using the KP formula~\eqref{eq:KPRed}.
For the small source $R=1~\text{fm}$, the characteristic dip structure is observed, which is typical in the system with a bound state below the threshold~\cite{Kamiya:2022thy}.
The difference among the results is clear in the low-momentum region of $q \le 100~\text{MeV/}c$.
The Isle potential model, which has the strongest repulsive core as seen in Fig.~\ref{fig:LamAlphaPot}, gives the most suppressed $C_{\Lambda\alpha}$. Through the comparison of different models, we find that the weaker the repulsion of the potential is, the less the suppression of the correlation function is.
On the other hand, this deviation is almost negligible in the results for $R = 3$ and $5~\text{fm}$. 
Remembering that the difference among the models mainly lies in their behavior at short range, while all the models reproduce the binding energy of ${}^5_\Lambda \mathrm{He}$,
this result suggests that the future measurement of the $\Lambda\alpha$ correlation function from a small source 
can constrain the $\Lambda N$ interaction at high densities.

\begin{figure*}[tbp]
\includegraphics[width=5.9cm]{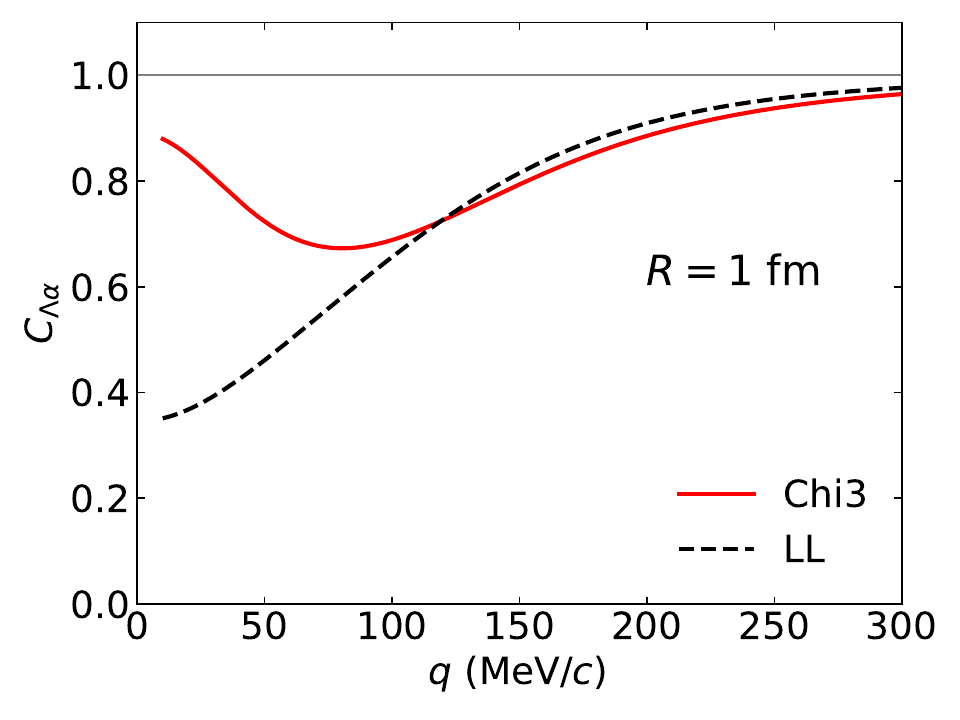}
\includegraphics[width=5.9cm]{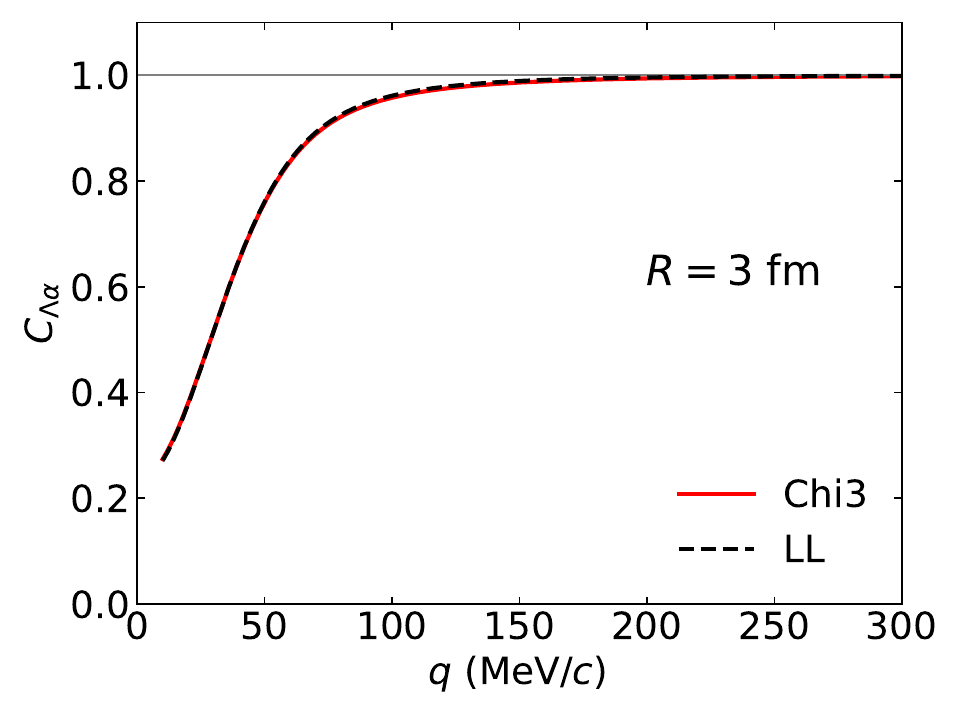}
\caption{$\Lambda\alpha$ correlation functions for two different source sizes.
The solid lines show the results calculated by the Chi3 $\Lambda\alpha$ potential with the KP formula.
The dashed lines represent the results using the LL formula
with the low-energy scattering parameters of Chi3.}
\label{fig:CF_LL}
\end{figure*}

In Fig.~\ref{fig:CF_LL}, the $\Lambda\alpha$ correlation functions calculated by using the KP formula with the Chi3 model
for different source sizes $R$ are compared with the ones from the LL formula~\eqref{eq:LL}
using the scattering length and the effective range of Chi3 in Table~\ref{tab:Properties}.
For $R=1~\text{fm}$, the LL model severely underestimates the results of the KP formula
in the $q \le 100~\text{MeV/}c$ region.
This reflects the fact that the LL formula cannot be applied to the case where the source size is smaller than the interaction range $\approx3~\text{fm}$.
This is a unique feature of the correlation functions including nuclei, because the interaction range of the hadron-hadron interactions is at most $1/m_{\pi}\sim 1.4$ fm.
For the larger source sizes, $R \ge 3~\text{fm}$,
the approximation by
the LL model works well.
This result may not be trivial because the use of the asymptotic wave function in the LL model is valid if $R$ is much larger than the interaction range, as mentioned above.

\begin{figure*}[tbp]
\includegraphics[width=5.9cm]{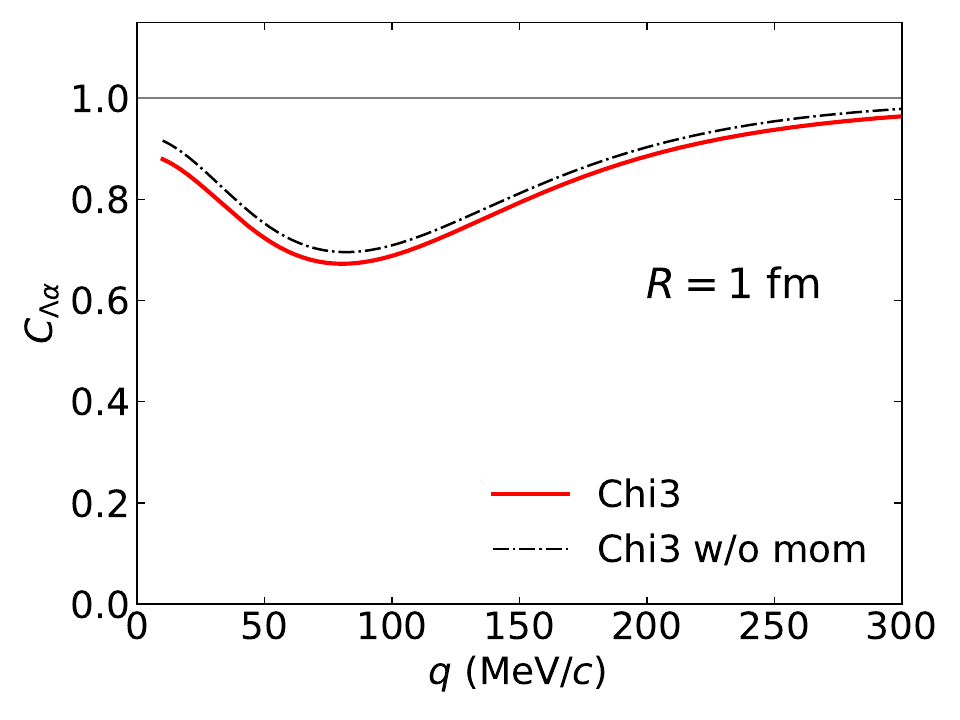}
\includegraphics[width=5.9cm]{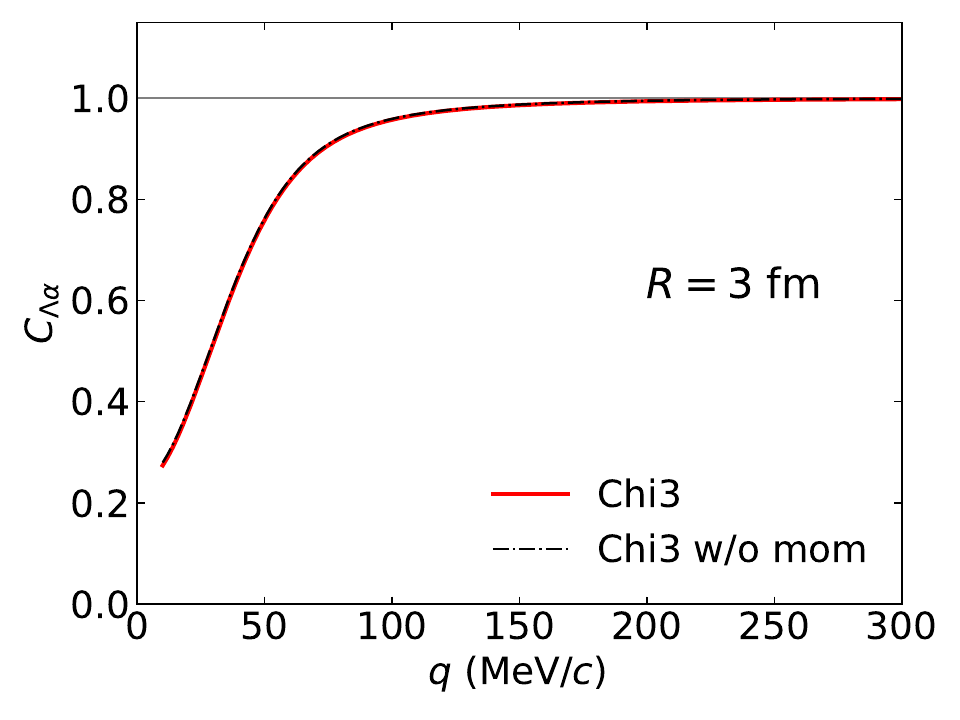}
\caption{$\Lambda\alpha$ correlation functions for different source sizes calculated by using Chi3 (solid line) and Chi3 w/o mom (dashed line).
}
\label{fig:CF_woMom}
\end{figure*}

To see the dependence on the momentum-dependent part of the Skyrme-type potential, we compare the $\Lambda\alpha$ correlation functions calculated by using Chi3 with those by Chi3 w/o mom in Fig.~\ref{fig:CF_woMom}.
For a source size of $R=1~\text{fm}$, the correlation functions show tiny but nonnegligible deviation, originated from the momentum dependence of the potential. From Fig.~\ref{fig:EffMass}, the momentum dependence of the potential induces a sizable difference in the reduced effective mass of the $\Lambda\alpha$ system. Nevertheless, its influence in the correlation function is quantitatively small, presumably because of the subsequent adjustment of the $a^\Lambda_3$ parameter 
to reproduce the $\Lambda$ binding energy of ${}^5_\Lambda \mathrm{He}$.
For $R\ge3~\text{fm}$, the differences in the correlation function are not noticeable.
For such larger source sizes, the LL formula works well, as seen above.
Then, the similarity between the correlation functions represents that the differences in $a_0$ and $r_{\rm eff}$
are not large enough to exhibit the difference in the correlation functions.

\section{summary}
\label{sec:summary}
In this paper, we extend the femtoscopy technique to the system including light nuclei, and we provide quantitative predictions of
the $\Lambda\alpha$ momentum correlation functions that can be measured in high-energy collisions.
We have examined five models of $\Lambda\alpha$ potentials.
Two of them are phenomenological $\Lambda\alpha$ models (Isle and SG)~\cite{Kumagai-Fuse:1994ulj}.
The others are constructed by substituting the $\alpha$ density distribution
for the Skyrme-type $\Lambda$ potentials~\cite{Jinno:2023xjr,Lanskoy:1997xq}.
All models reproduce the $\Lambda$ binding energy of $^5_\Lambda \mathrm{He}$ and have a consistent
interaction range of $2$-$3~\text{fm}$,
while they have different properties at short range,
including both attractive ones and repulsive ones.
The constructed Skyrme-type potentials indicate that the repulsive nature of the $\Lambda$ potential at high densities induces the repulsive core in the $\Lambda\alpha$ interaction at short range.

While the correlation functions from the source with $R\gtrsim 3$ fm are not sensitive to the short-range behavior of the $\Lambda\alpha$ potential, the difference of the potentials is manifest in the correlation functions from the small-source system ($R\sim 1$ fm). It is found that the correlation is
suppressed in the order of the repulsive strength of the $\Lambda\alpha$ potential at short range.
This indicates that the $\Lambda\alpha$ correlation function can constrain
the $\Lambda\alpha$ potential at short range, which cannot be explored in the few-body $\Lambda$ hypernuclear system because the variation in its short range part does not make a difference in the calculated $\Lambda$ binding energy~\cite{Motoba:1985uj}.
Detailed knowledge of the $\Lambda\alpha$ potential at short range would
provide valuable information on the property of $\Lambda$ in dense nuclear medium, which is one of the key ingredients needed to solve the hyperon puzzle of neutron stars.

We examine the validity of the LL formula, which has been utilized
to extract the low-energy scattering parameters from the correlation function measurements.
For a small source size of $1~\text{fm}$,
the LL formula is shown to severely deviate from the exact result in the low-momentum region,
since the system with longer interaction range than the source size invalidates the assumption made in the LL formula.
We also study the effect of the momentum dependence of the $\Lambda$ potential,
which is not so firmly determined from the experimental data.
We compare the momentum dependent model with the one omitting the momentum dependence of the $\Lambda$ potential in symmetric nuclear matter while fixing the $\Lambda$ binding energy of ${}^5_\Lambda\mathrm{He}$.
The difference between with and without the momentum dependence
is found to be small.

Since the source size of $1~\text{fm}$ is smaller than the rms radii of $\alpha$,
the feasibility of treating $\alpha$ as a point-like particle should be discussed.
In the coalescence model picture, the yield of the composite particle is represented as the product of the single-particle yields and their correlation, and then the source function of the composite particle can be regarded as the effective Gaussian source function~\cite{Mrowczynski:2019yrr,Mrowczynski:2020ugu,Bazak:2020wjn,Mrowczynski:2021bzy}.
A more rigorous treatment for treating the $\alpha$ particle as a composite particle is to calculate the five-body scattering problem of $\Lambda + 2n + 2p \rightarrow \Lambda + \alpha$.
However, performing such calculations is beyond the scope of this paper and is left as a future work.

We have demonstrated that the study of the two-body correlation functions including $\alpha$ could serve as a new tool to study the property of the hyperons in nuclear medium.
The experimental measurement of the $\Lambda\alpha$ correlation function
may be feasible at the collision energy $\sqrt{s_{NN}}<10~\text{GeV}$ in which a number of $\alpha$ particles would be produced in central heavy-ion collisions as estimated by the statistical model~\cite{Andronic:2010qu}.
Also, according to Ref.~\cite{Andronic:2010qu}, the yield of $\Lambda$ is always larger than that of $\alpha$ for $\sqrt{s_{NN}}\ge3~\text{GeV}$.
We hope that the present work stimulates the study of the $\Lambda\alpha$ correlation functions in future experiments,
including the facilities with medium-collision energies such as FAIR~\cite{CBM:2016kpk}, NICA, and J-PARC HI~\cite{Ozawa:2022sam}.

\begin{acknowledgments}
We thank Kouichi Hagino, Yudai Ichikawa, Johann Haidenbauer, Andreas Nogga, Hoai Le, and Stanis\l{}aw Mr\'owczy\'nski for useful discussions and comments.
This work was supported in part by Grants-in-Aid for Scientific Research from JSPS
(Grants 
No.~JP23H05439 and % Kiban S (Hyodo)
No.~JP22K03637), % Kiban C (Hyodo)
by JST, the establishment of university fellowships towards
the creation of science technology innovation,
Grant No.~JPMJFS2123, and
by the Deutsche Forschungsgemeinschaft (DFG) and the National Natural Science Foundation of China (NSFC) through the funds provided to the Sino-German Collaborative Research Center ``Symmetries and the Emergence of Structure in QCD" (NSFC Grant No. 12070131001 and DFG Project-ID 196253076 -- TRR 110).
\end{acknowledgments}

\bibliography{ref,ref_LamBE,ref_Femto}

\end{document}